\documentclass{article}%
\usepackage{amsmath}%
\usepackage{amsfonts}%
\usepackage{amssymb}%
\usepackage{graphicx}

\begin{document}

\title{ A new method of measurement of the velocities of solar neutrinos}
\author{Elmir Dermendjiev\\"Mladost-2", block 224, entr.1, apt.13, 1799-Sofia, Bulgaria}
\maketitle

\begin{abstract}
A new method of measurement of the velocities of solar electron antineutrinos
is proposed. The method is based on the assumption, that if the neutrino
detector having a shape of a pipe and providing a proper angular resolution,
is directed onto the optical "image" of the sun, then it would detect solar
neutrinos with velocities $V_{\widetilde{\nu}_{e}}$ $=c$. Here c is the
velocity of light. It is expected that the less is the value of $V_{\widetilde
{\nu}_{e}}$, the larger would be the angular lagging of the "image" of these
neutrinos relative to the position of the optical "image" of the sun.
Therefore, one can detect solar neutrinos with different energies by changing
the angle between the axis of a detector pipe and the direction to the "image"
of the sun. Also, the method gives unique possibility to check hypotheses
predicting an existence of solar neutrinos with $V_{\widetilde{\nu}_{e}}>c$.
In this case the "image" of such solar neutrinos on the sky should pass the
"image" of the sun.

\end{abstract}


Solar electron antineutrino; Velocities of solar

electron antineutrinos; method of measurement of velocities of solar

electron antineutrinos.

\section{ Introduction}

The problem of a generation of the solar energy and its nature is probably the
most important astrophysical one. At present the most spread notion on the
origin of the solar energy is based on the assumption that it appears as a
result of thermonuclear reactions inside the sun. Briefly, the process of
transformation of hydrogen to helium is accompanied by emission of electron
antineutrinos $\widetilde{\nu}_{e}$ [1]. Some astrophysical solar models
predict different yields of groups of antineutrinos with different energies
$E_{\nu}$. In regard with the "Standard Solar Model" (SSM) most of
antineutrinos ($\symbol{126}99.75\%$) should have an energy in the range of 0
$<$
$E_{\nu}$
$<$
$420keV$ (so - called "PP1" - group of neutrinos). The group of "PeP" -
antineutrinos has a fixed energy $E_{\nu}=1.44MeV$, but much lower intensity
than the first group. A group "HeP" has an energy spectrum up to $18.6MeV$ and
the yield of $\symbol{126}10^{-5}\%$, etc. In addition, the theory predicts
the existence of two other groups of antineutrinos:"PP2" and "PP3". They have
mono-energetic and continuous energy spectra respectively and are important
for the theory. Other source of antineutrinos is so-called "carbon-nitrogen"
(CN) cycle, that generates two more groups of antineutrinos with $E_{\nu
}=1.2MeV$ and $1.7MeV$. Thus, to check the SSM one needs to perform very
complicated experiment that includes a spectrometry of the energies of solar
antineutrinos and a measurement of the intensities of these groups of
$\widetilde{\nu}_{e}$ . Unfortunately, at present such experiment is out of
our technical possibilities due to many reasons. The most serious experimental
difficulty in performing astrophysical experiments with solar $\widetilde{\nu
}_{e}$ is an extremely low absorption cross section $\sigma_{a}$%
($\widetilde{\nu}_{e}$ ) of electron antineutrinos by nuclei, which is
estimated to be of $\symbol{126}10^{-43}cm^{2}$[2].

At present solar $\widetilde{\nu}_{e}$ are studied in a few laboratories [1,3]
by using giant neutrino detectors, which are briefly discussed below.
Unfortunately, these facilities are not intended for detection of the
relatively low energy solar antineutrinos or their values of $V_{\widetilde
{\nu}_{e}}$.

Below, a method that might be suitable for measurements of the velocities
$V_{\widetilde{\nu}_{e}}<c$ of solar electron antineutrinos is proposed. One
can hope that the experimental information obtained by proposed method might
be useful for further development of the SSM.

At present, the proposed method of measurement of the velocities of solar
antineutrinos is the only method, that allows to search whether some of them
have values of $V_{\widetilde{\nu}_{e}}>c$. Possible existence of such
hypothetical solar antineutrinos is discussed in [4]. This hypothesis is
inconsistent with the theory of relativity. However, by using the method
proposed below, one has the unique possibility to find such super-fast
particles with $V_{\widetilde{\nu}_{e}}>c$, if they exist, or to refute the
hypothesis [4].

\section{ A method of measurement of the velocities of solar antineutrinos}

The proposed method of measurement of the velocities $V_{\widetilde{\nu}_{e}}$
of solar electron antineutrinos is based on the simple and clear idea that the
position of their "images" on the sky relative to the optical "image" of the
sun should depend on the value of $V_{\widetilde{\nu}_{e}}$. Assume, that the
observer has a $"pipe"$-type neutrino detector that follows the movement of
the optical "image" of the sun on its sky trajectory. Also, assume that this
detector has an angular resolution $\Delta\alpha$ comparable with the angular
size of the sun, i.e. $\Delta\alpha\leq0,5^{o}$. Then, if $V_{\widetilde{\nu
}_{e}}=c$, both the neutrino and the sun "images" should coincide. In the case
when $V_{\widetilde{\nu}_{e}}$ $<c$, the "image" of given group of neutrinos
with certain value of $V_{\widetilde{\nu}_{e}}$ is expected to have an angular
lagging relative to the optical "image" of the sun. It is clear that the less
is the value of $V_{\widetilde{\nu}_{e}}$, the larger would be the angular
lagging of $\widetilde{\nu}_{e}$  having that value of $V_{\widetilde{\nu}%
_{e}}$ . In the case when $V_{\widetilde{\nu}_{e}}>c$, one should expect that
the "image" of such neutrinos should pass the "image" of the sun. Thus, if the
neutrino detector follows the "image" of the sun with a constant angle $\beta$
of lagging or passing, then one can determine the value of $V_{\widetilde{\nu
}_{e}}$ .

How large are the expected values of $\beta$? The angular velocity $\omega$ of
the optical "image" of the sun relative to the Earth is of $\omega
=0.00417deg.s^{-1}$. Two cases are discussed below:

a) $V_{\widetilde{\nu}_{e}}<c$. If $\beta$ does not exceed a few degrees, then
the value of $V_{\widetilde{\nu}_{e}}$

can be estimated by using the following approximate relationship:%
\begin{equation}
V_{\widetilde{\nu}_{e}}\thickapprox\frac{L}{480+\beta/\omega}%
\end{equation}

Here $L=1.45.10^{11}m$. If, for instance, $V_{\widetilde{\nu}_{e}}%
=10^{8}m.s^{-1}$ or $3.10^{7}m.s^{-1}$,

then the value of $\beta$ is approximately $4^{o}$ or $18^{o}$.

b) $V_{\widetilde{\nu}_{e}}$ $>c$. In this case the "image" of such super-fast
hypothetical neutrinos should

pass the optical "image" of the sun with an angle $\beta\prime$. Then a similar

relationship can be used:%
\begin{equation}
V_{\widetilde{\nu}_{e}}\thickapprox\frac{L}{480-\beta^{\prime}/\omega}%
\end{equation}

It is interesting to note that if $V_{\widetilde{\nu}_{e}}\rightarrow$
$\infty$, then the maximum value of

$\beta^{\prime}$ is close to $2^{o}$, i.e. of $\symbol{126}\ 4$ angular sizes
of the diameter of the

sun. This means, that if such super-fast solar neutrinos would

exist, than they could be found experimentally.

\section{\bigskip\ Discussion}

The study of the properties of solar neutrinos in modern laboratories [1,3] is
an extremely difficult task. As it was briefly mentioned above, there are some
experimental difficulties originated by very low value of the absorption cross
section $\sigma_{a}$($\widetilde{\nu}_{e}$) of solar antineutrinos by nuclei
[2]. This leads to a lack in detection rate even in the case of using huge
solar neutrino detectors. Another difficult problem is the necessity the
detector background to be minimized. It requires neutrino detectors to be
situated underground.

To conclude whether the proposed method could be applied in solar neutrino
experiments, brief comparison between some of the existing methods of
detection of solar neutrinos and the proposed method is presented below.

Up to now, there are only a few experimental studies of solar neutrinos, which
are not considered in this paper. However, all of them were performed with
giant neutrino detectors, like the solar neutrino detector designed by Davis
[1]. This large facility has 615 tons of $C_{2}Cl_{4}$, used as a detector
substance. The detection of solar antineutrino is based on the reaction of
absorption of $\widetilde{\nu}_{e}$ by a nucleus of $^{37}Cl$ [5]:%
\begin{equation}
^{37}Cl+\widetilde{\nu}_{e}\rightarrow^{37}Ar+e^{-}%
\end{equation}
This reaction has a threshold of $0.816MeV.$ The detection of solar
antineutrinos is based on the radiochemical reaction (3) and the
detector cannot be used for energy measurements. The SNO [4] is
another large facility that contains 1000 tons of $D_{2}O$ for study
of the properties of solar neutrinos. The detection of
$\widetilde{\nu}_{e}$ is released
when $\widetilde{\nu}_{e}$ interacts with deuterium nuclei:%
\begin{equation}
d+\widetilde{\nu}_{e}\rightarrow p+p+e^{-}%
\end{equation}

The SNO facility is intended to be used manly for detection of high
energy neutrinos. Therefore, it is not suitable for spectrometry of
energies or velocities of solar antineutrinos. On the other hand,
further development of the SSM needs the intensities of different
groups of antineutrinos with fixed energies
$E_{\widetilde{\nu}_{e}}$ to be known with permanently arising
accuracy. Since there are no neutrino detectors capable to measure
the energies of solar neutrinos, one can hope that the proposed
method of measurement of the velocities of neutrinos could, to some
extend, contribute the solution of this very important problem of
generation of the solar energy. The counting rate
$N_{\widetilde{\nu}_{e}}$ of a "$pipe$"-type neutrino detector is
estimated below. It is desirable the value of $\Delta\alpha$
$\symbol{126}(\Phi/L)$ of this detector to be comparable with the
angular size of the optical "image" of the sun, which is of
$\symbol{126}0.5^{o}$. Also, it is necessary the value of
$\Delta\alpha\symbol{126}0.5^{o}$ to be kept, if one would search
for hypothetical "super-fast" neutrinos [4]. Suppose that the length
$L$ and the diameter $\Phi$ of the $pipe$ are of $11.5m$ and $0.1m$
respectively. It seems to be reasonable, a detection technique,
similar to that described by Reines at al. [2], to be chosen. The
$pipe$ contains liquid scintillator with small amount of Cd (or Gd).
This $pipe$ is inserted into the outer $pipe$ with larger diameter
and six radial sections along the pipe. All radial sections are
filled with liquid scintillator. Photomultiplier tubes are mounted
along the section sides of outer $pipe$. The detection of solar
electron antineutrino is based on the following nuclear reaction:%
\begin{equation}
\widetilde{\nu}_{e}+p\rightarrow n+e^{+}%
\end{equation}
Two annihilation gamma-quanta mark the reaction (5) and \symbol{126}
six slow neutron capture gamma-quanta are emitted due to the
$Cd(n,\gamma)$ reaction [2]. A delayed coincidence technique used in
[2] provides quite low detection background. Similar technique was
successfully used in fission [6] and sub-threshold fission
experiments [7] with $U$, $Pu$ and $Np$ isotope targets having very
high specific $\alpha$ - and $\gamma$ - activity. \qquad\ The
counting rate $N_{\widetilde{\nu}_{e}}$ can be estimated by using
the approximate relationship:
\begin{equation}
N_{\widetilde{\nu}_{e}}\thickapprox n\sigma_{a}(\widetilde{\nu}_{e}%
,p)\phi\epsilon_{\widetilde{\nu}_{e}}SLt
\end{equation}
Here, $n$ is the number of hydrogen atoms in$1cm^{3}$ of liquid
scintillator,
$\sigma_{a}(\widetilde{\nu}_{e},p)=1.2.10^{-43}cm^{2}$[2]. The
approximate value of the flux of solar electron antineutrinos at the
surface of the Earth is accepted to be of
$\phi\thickapprox5.10^{10}cm^{-2}s^{-1}$. The detection efficiency
$\epsilon_{\widetilde{\nu}_{e}}$ is estimated to be of
$\symbol{126}0.3$ if a delayed coincidences between two annihilation
gamma-quanta and more than three captured gamma-quanta are realized.
In Eq.(6) $\ S=(\pi/4)\Phi^{2}$, were $\Phi=0.1m$, $L=11.5m$ and$\
t=1s.$
Using these numbers one gets an estimated value of $N_{\widetilde{\nu}_{e}%
}\symbol{126}10^{-4}s^{-1}$, which means that one could collect
$\symbol{126}3.10^{3}$ events per year. Further optimization of the
"pipe"-type neutrino detector might strongly reduce the measurement
time compare to the estimated value. Brief consideration of
different methods of detection of electron antineutrinos, allows to
be concluded, that the proposed method could be used for
measurements of velocities of solar antineutrinos and, thus, provide
a new experimental data for further development of the SSM. Also,
having a moderate size, the proposed "$pipe$"-type neutrino detector
can ensure a reasonable time of measurements. However, the most
important preference of proposed method is the opportunity to search
for existence of neutrinos with $V_{\widetilde{\nu}_{e}}>c$. Based
on the theory of relativity, one should not expect an existence of
such neutrinos in Nature. Then, the experiment based on the proposed
method of measurement the velocities of neutrinos would be a strong
confirmation of the theory of relativity. But, if neutrinos with
$V_{\widetilde{\nu}_{e}}>c$ exist, one should expect deep change of
our understanding of Nature.

\end{document}